# PHYSICALLY DETACHED "COMPACT GROUPS"

LARS HERNQUIST[1], NEAL KATZ[2] & DAVID H. WEINBERG[3]

## ABSTRACT

A small fraction of galaxies appear to reside in dense compact groups, whose inferred crossing times are much shorter than a Hubble time. These short crossing times have led to considerable disagreement among researchers attempting to deduce the dynamical state of these systems. In this paper, we suggest that many of the observed groups are not physically bound but are chance projections of galaxies well-separated along the line of sight. Unlike earlier similar proposals, ours does not require that the galaxies in the compact group be members of a more diffuse, but physically bound entity.

The probability of physically separated galaxies projecting into an apparent compact group is non-negligible if most galaxies are distributed, as observed, in thin filaments. We illustrate this general point with a specific example: a simulation of a cold dark matter universe, in which hydrodynamic effects are included to identify galaxies. The simulated galaxy distribution is filamentary, and end-on views of these filaments produce apparent galaxy associations that have sizes and velocity dispersions similar to those of observed compact groups. The frequency of such projections is sufficient, in principle, to explain the observed space-density of groups in the Hickson catalog. We discuss the implications of our proposal for the formation and evolution of groups and elliptical galaxies. The proposal can be tested by using redshift-independent distance estimators to measure the line-of-sight spatial extent of nearby compact groups.

[1] Sloan Foundation Fellow, Presidential Faculty Fellow; Lick Observatory, U.C. Santa Cruz, Santa Cruz, California 95064
[2] Department of Astronomy, University of Washington, Seattle, Washington 98195
[3] Institute for Advanced Study, Princeton, New Jersey 08540

## 1. INTRODUCTION

The first known example of a small group of galaxies in close proximity was that discovered by Stephan more than a century ago (Stephan 1877). Although the second system of this type was not reported until more than seventy years had elapsed (Seyfert 1948, 1951), many more have been found in the past twenty years, at least partly in response to the recognition that interacting *pairs* of galaxies are not uncommon (Vorontsov-Velyaminov 1959; Arp 1966). Careful searches have revealed $\sim 100$ such "compact" groups containing four or more relatively large galaxies (Rose 1977; Hickson 1982, 1993). Indeed, it has been estimated that compact groups contain roughly one percent of all visible matter (Mendes de Oliveira & Hickson 1991).

The physical state of these systems is controversial. Based on projected separations and line-of-sight velocity differences, it is estimated that they have typical crossing times of order one percent the age of the Universe and mass-to-light ratios $M/L \sim 50h$, where $h \equiv H_0/100$ (Hickson *et al.* 1992; Hickson 1993). Theoretical arguments and computer simulations have shown convincingly that physically bound compact groups would experience a "merging" instability, which would lead to their complete coalescence within several dynamical times (Hickson *et al.* 1977; Carnevali, Cavaliere & Santangelo 1981; Ishizawa *et al.* 1983; Ishizawa 1986; Mamon 1987; Barnes 1984, 1989). While the theoretically estimated lifetime of a compact group may be lengthened by evolutionary effects (Barnes 1985), by the possibility that the galaxies in question share a common dark-matter halo (see, *e.g.* Barnes 1985; Navarro, Mosconi & Garcia-Lambas 1987), or by a spectrum of galaxy masses (*e.g.* Governato, Bhatia & Chincarini 1991), recent estimates suggest that most observed groups should merge in a time much shorter than a Hubble time (for discussions, see, *e.g.* White 1990; Puche & Carignan 1991; Athanassoula & Makino 1994; Bode, Cohn & Lugger 1994). The short lifetimes are puzzling unless, as emphasized by White (1990), the progenitors and descendents of observed compact groups can be identified.

Within the framework of hierarchical models for structure formation, a natural interpretation of compact groups is that they are, in fact, transient, and that they quickly merge to form giant elliptical galaxies (Barnes 1985, 1989). Their population is continuously replenished, however, as loose aggregates within more diffuse groups become tightly bound and collapse (*e.g.* White 1976; Cavaliere *et al.* 1986). The statistics of this scenario seem roughly consistent with observations (Nolthenius & White 1987; Barnes 1989), although it has yet to be demonstrated that the needed rates arise from plausible initial conditions. For example, the formation rate of ellipticals and compact groups in this model will depend on cosmological parameters such as $\Omega$ and $\Lambda$, and some fine-tuning may be required to reproduce the number of compact groups known at the present time. A detailed examination of the loose group $\to$ dense group $\to$ merger scenario has recently been carried out by Diaferio, Geller & Ramella (1994) and Ramella *et al.* (1994).

The difficulties of short dynamical timescales have led some authors to suggest that most of the observed compact groups are not physically bound but arise instead from projection effects. One possibility is that compact groups are transient projections of galaxies bound only indirectly to one another through their common membership in loose groups (*e.g.* Rose 1979; Mamon 1986, 1987; Walke & Mamon 1989). While this proposal may be correct in certain cases, quantitative evidence in its favor is lacking because the various attempts to study it have generally ignored the effects of dissipation and/or the self-gravity of the galaxies, and have not employed models derived from underlying cosmological initial conditions. In addition, it does not appear that loose groups are sufficiently dense to account for the number of compact groups seen (Hickson & Rood 1988).

In this paper, we propose an alternative interpretation of compact groups; namely, that they are chance projections of galaxies that need not even be bound to one another indirectly. Galaxy redshift surveys show that a large fraction of galaxies lie in thin sheets and filaments (*e.g.* de Lapparent, Geller & Huchra 1986; Haynes & Giovanelli 1986; Geller & Huchra 1989). Filamentary structure arises naturally in theories of gravitational instability with Gaussian initial conditions (Shandarin & Zel'dovich 1989), and it is seen in numerical simulations of structure formation (*e.g.* White *et al.* 1987; Melott & Shandarin 1993). As we show in §3, if the galaxy distribution is filamentary, then the probability of viewing an unbound collection of galaxies in close projected proximity is non-negligible.

For a specific illustration of this idea, we analyze a hydrodynamic simulation of structure formation in a cold dark matter (CDM) universe. Incorporating dissipative hydrodynamic effects in the simulation allows us to identify galaxies in a robust and straightforward way. Chance projections of filamentary features produce galaxy associations with structure, kinematics, and frequency similar to those of observed compact groups. While we examine only the CDM model in detail — and even in this case our statistical accuracy is limited by our small simulation volume — we believe that our suggestion is fairly model-independent, provided that galaxies are arranged along filaments. In addition to eliminating the puzzle of short dynamical times, our proposal can naturally explain the linear appearance of some compact groups.

In §2, we describe technical details of our simulation. Results relevant to the physical nature of compact groups are summarized in §3. Finally, possible implications and observational tests are outlined in §§4 & 5.

## 2. METHODOLOGY

The calculations presented here are part of an ongoing study to examine galaxy formation in various cosmological models using numerical simulation. We use a hybrid N-body/hydrodynamics code (Hernquist & Katz 1989; Katz, Weinberg, & Hernquist 1994) to follow the simultaneous evolution of collisionless matter and gas in an expanding universe. Specifically, our code uses smoothed particle hydrodynamics (SPH; Lucy 1977; Gingold & Monaghan 1977) to integrate the fluid equations which describe the gas, and a hierarchical tree code to compute gravitational forces (Barnes & Hut 1986). In all cases, the simulation volume is taken to be a periodic cube (Bouchet & Hernquist 1988; Hernquist, Bouchet & Suto 1991).

Some preliminary results from our effort have been described elsewhere (Katz, Hernquist & Weinberg 1992). Briefly, we consider the growth of structure in a conventional CDM model with $\Omega = 1$ and $H_0 = 50$ km s$^{-1}$ Mpc$^{-1}$. The power spectrum is normalized to an rms linear fluctuation of 0.7 in spheres of radius 16 Mpc; in common terminology, a "bias" of $1/0.7 = 1.44$. In §3, we report results from a simulation including gas in which the baryon fraction is $\Omega_b = 0.05$. This model is similar to that discussed by Katz *et al.* (1992), but it has been evolved to a redshift $z = 0$, at which time we search this 22 Mpc cube for fraudulent groups. (All distances, unless otherwise specified, are for $H_0 = 50$.)

The small-scale resolution of our simulations is limited by the gravitational softening length and by the SPH smoothing length. In what follows, the gravitational softening length is fixed at 20 kpc (comoving) for all particles, and each gas particle has its own smoothing length, which is continuously updated so that 32 neighbor particles are within each smoothing volume. Dark matter particles have a constant time-step of $t_0/4000$, where $t_0 = 2H_0/3$, while gas particles have variable time-steps chosen to satisfy a modified form of the Courant condition. The latter ensures that all the gas particles have time-steps at least as small as that of the dark matter particles. The dark matter particle mass is $2.20 \times 10^{10} M_\odot$, and the gas particle mass is $1.16 \times 10^9 M_\odot$.

It should be stressed that we have chosen to investigate the hypothesis that many observed compact groups are chance superpositions using a hydrodynamic simulation not because the baryonic matter alters the large-scale mass distribution but because it is trivial to identify galaxies in such a simulation from the location of dense knots of cold gas (Katz *et al.* 1992; Evrard, Summers & Davis 1994). Unfortunately, we have only one such simulation evolved to zero redshift at our disposal, and the volume of the simulation is rather small. We have also examined the spatial distributions of halos in collisionless N-body simulations, for which we find qualitatively similar results. Studies that identify galaxies with high peaks of the initial density field might underestimate the projection effects found here because "peaks biasing" tends to overpopulate real (*i.e.* 3-dimensional) clusters and to underpopulate filaments (Katz, Quinn & Gelb 1993).

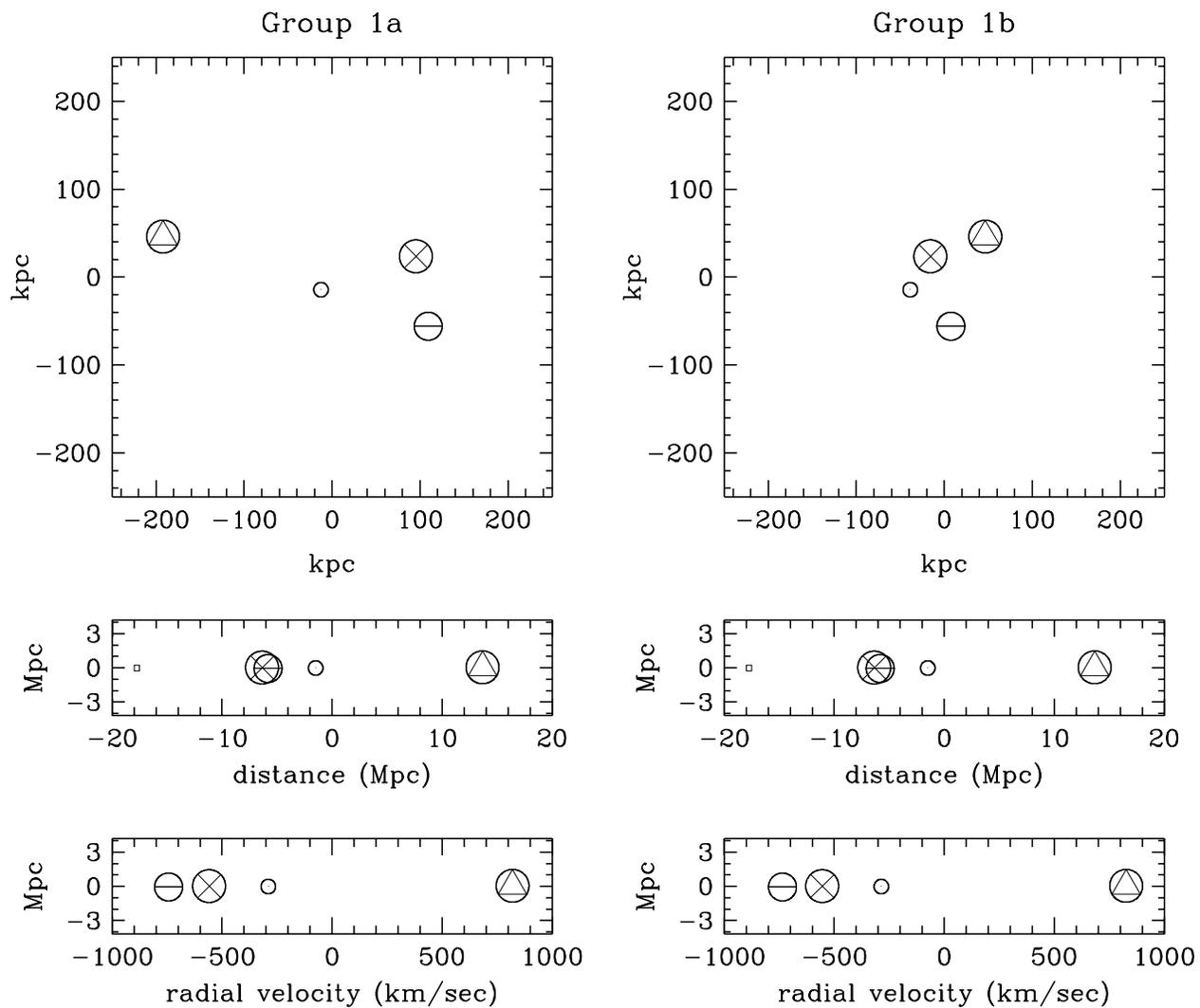

**Figure 1** — A montage of false "compact groups" identified by viewing filaments along their lengths in a cosmological simulation of structure formation in a CDM universe. To identify groups, we employ the selection criteria adopted by Hickson (1993). Each frame illustrates the structure of a false group in our simulation. The top panel shows the projected spatial distribution of the galaxies, the middle panel shows the relative distances of these galaxies along the line of sight, plotted against the vertical coordinate from the top frame, and the bottom panel plots the radial velocity (Hubble flow plus peculiar motion) against the same vertical coordinate. In all panels the vertical scale is chosen to match the horizontal scale, and the origin is located at the group's center of mass. The small square on the left side of the middle panel shows, to scale, the $(500\ \text{kpc})^2$ area of the top panel. The symbol identifying the location and velocity of a galaxy is proportional in area to the galaxy's mass. Different filaments are labeled 1 through 5. Letters associated with a group number indicate slight variations in viewing angle.

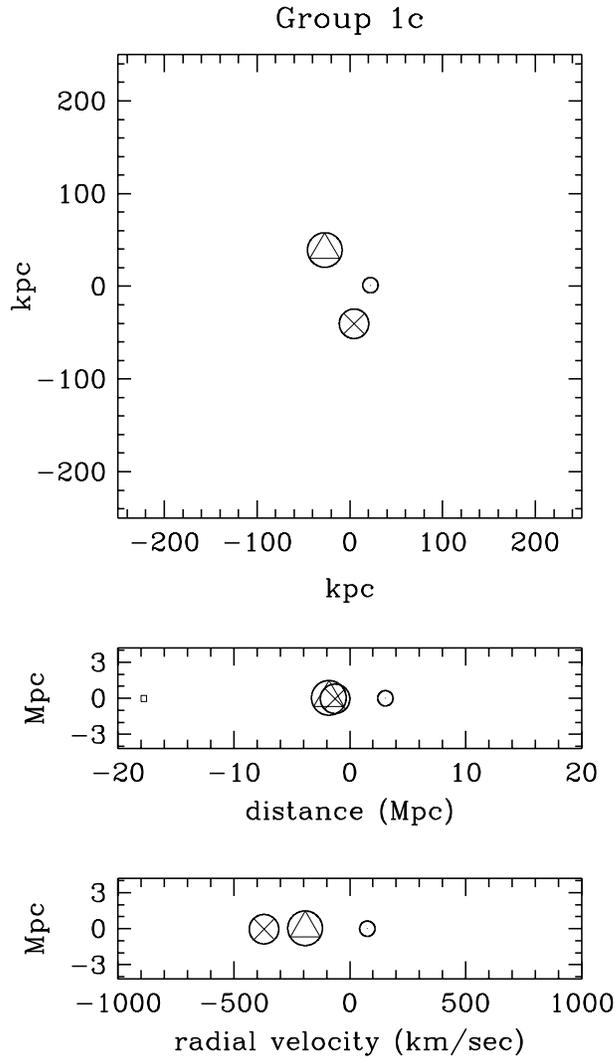

**Figure 1** — continued.

## 3. RESULTS

Figure 1 displays a collection of "compact groups" identified in projection from the galaxy distribution of the CDM simulation described in §2. We select groups according to the criteria of Hickson's (1993) catalog: a group must be fairly isolated, it must have at least four members, and its member galaxies must be within three magnitudes of one another in luminosity. Our simulations do not provide us with direct information on the luminosities of individual galaxies, so we assume that all galaxies have the same *baryonic* mass-to-light ratio in order to convert their gas masses into luminosities. One of the groups in Figure 1 contains only three galaxies; since many of Hickson's four-member groups include at least one object with a discordant redshift, it seems reasonable to include one three-member system from this small simulation cube in our "catalog."

The frames in Figure 1 show projections along five separate filaments in the simulation volume. For each, we illustrate how slight variations in viewing angle can affect the group's projected morphology and line-of-sight velocity structure. Thus, Groups 1a, 1b, and 1c refer to different projections along the *same* filament, while Groups 1 and 2, for example, refer to projections along *different* filaments. The top panel in each frame shows the projected spatial distribution of the galaxies. The middle panel in each frame shows the relative distances of these galaxies along the line of sight, plotted against the vertical coordinate in the top panel. The bottom panel shows the radial velocity of each galaxy plotted versus the vertical coordinate in the top panel; this velocity includes both Hubble flow and peculiar motion. In every panel the vertical scale is chosen to match the horizontal scale. The symbols identifying the locations of the galaxies are proportional in area to the galaxy masses.

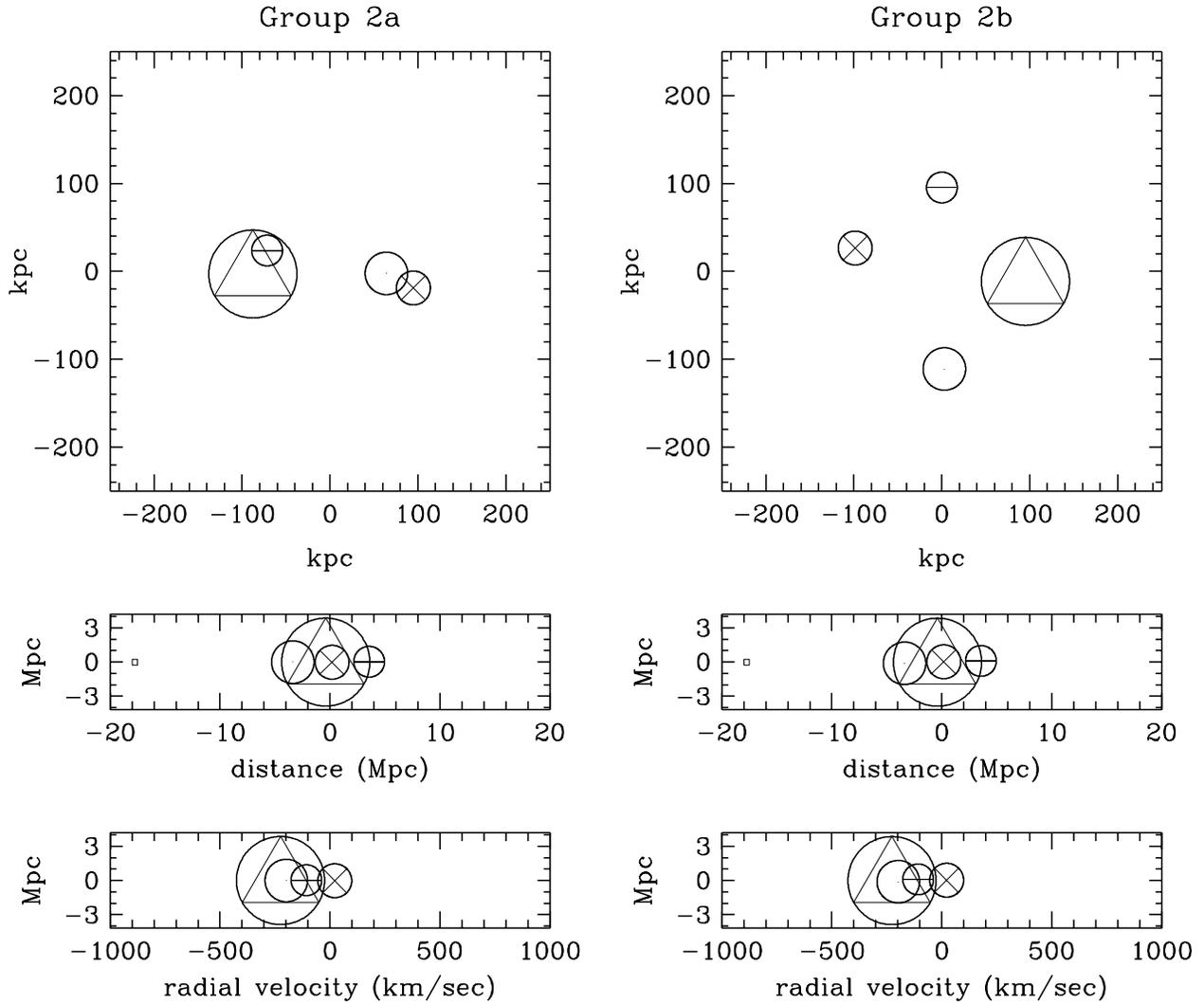

**Figure 1** — continued.

In terms of their appearance projected onto the plane of the sky and the relative velocities of the galaxies they contain, the systems depicted in Figure 1 closely resemble those in Hickson's (1993) atlas. Of course, as the middle panel in each frame demonstrates, the galaxies in each of our groups are widely separated in three-dimensional space, with typical radial extents of many Mpc. Groups 1a and 1b contain one physically associated pair and two projected galaxies. The physical extent of this system is nearly 20 Mpc. Group 1c adopts a somewhat different viewing angle and therefore loses one of the projected galaxies. Groups 2a and 2b have one close pair and two other members 2-4 Mpc away. Their line-of-sight velocity dispersions are below 100 km/s despite radial extents of nearly 7 Mpc. Groups 3a and 3b are similar to 2a and 2b, but with somewhat higher velocity dispersions. Groups 4a, 4b, and 4c consist of one associated pair and a projected string of three small galaxies. Groups 5a and 5b contain four low mass galaxies, none of which form close pairs in 3 dimensions.

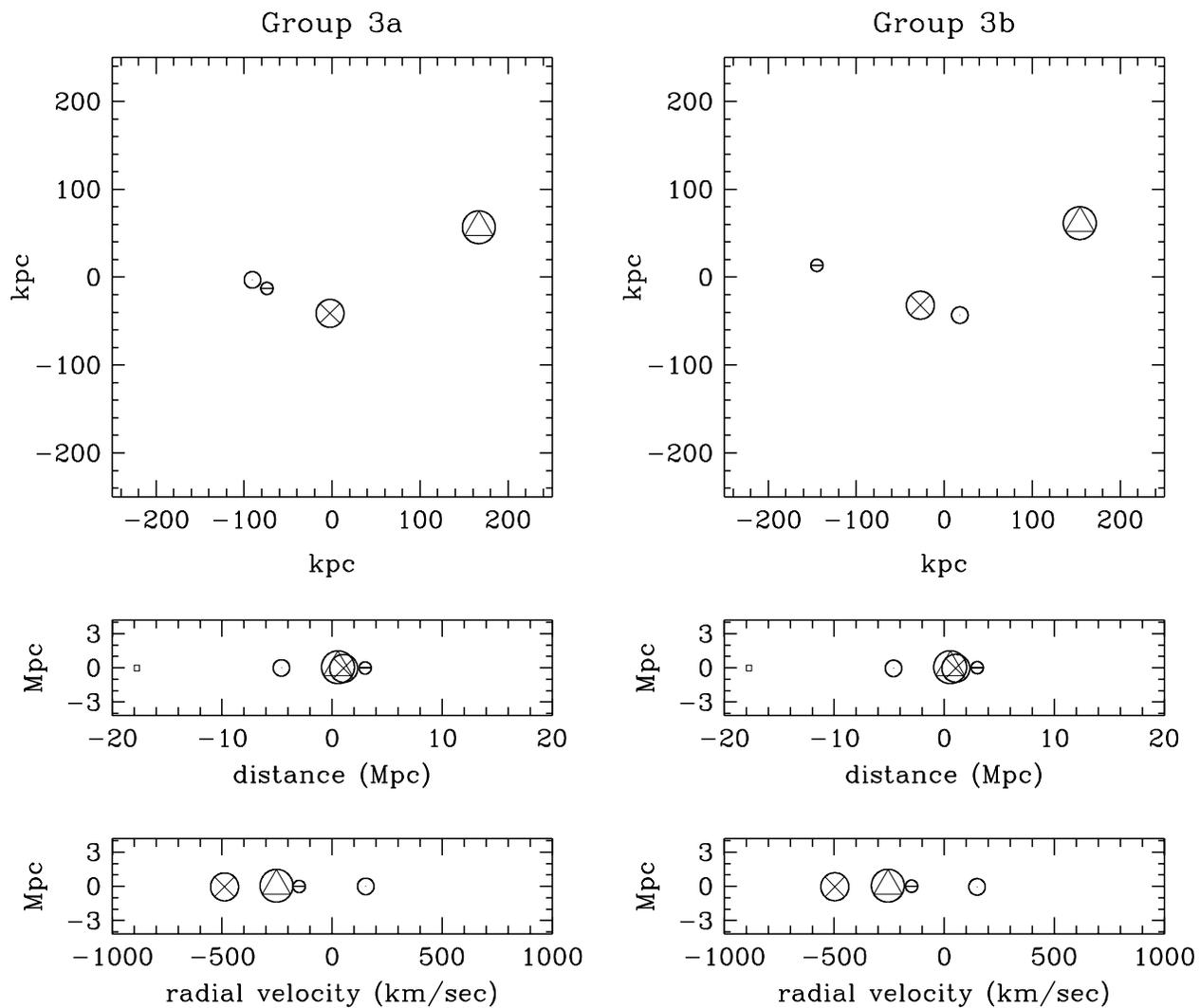

**Figure 1** — continued.

Table 1 lists properties of the twelve groups depicted in Figure 1. In particular, the fact that the groups are physically extended, despite their small projected sizes, is verified quantitatively in the sixth column, which gives the full extent of each group along the line of sight, $\Delta z$. The second column of Table 1 gives the projected size of each group, $R$, where $R$ is computed by taking the median separation between galaxy pairs, as in Mendes de Oliveira (1992) and Hickson et al. (1992). In every case, $\Delta z \gg R$, owing to the filamentary nature of the galaxy distribution in CDM models.

To compare our results with the dynamical estimates of Hickson et al. (1992) and Hickson (1993), we estimate group sizes, $R$, as described above, and three-dimensional velocity dispersions, $\sigma_{3D}$, as if the groups were indeed bound. Following Mendes de Oliveira (1992) and Hickson et al. (1992), we compute the line of sight velocity dispersion for each group, $\sigma_z$ from $\sigma_z^2 = <v^2> - <v>^2$, where $<>$ denotes an average of the radial velocities for galaxies in each group, and we use this to estimate $\sigma_{3D}$ from $\sigma_{3D}^2 = 3\sigma_z^2$. Values of $\sigma_z$ are given in column 3 of Table 1. A crossing time, $t_c$ can be obtained from these quantities through $t_c = 4R/\pi\sigma_{3D}$. The crossing times for our groups are listed in columns 4 and 5 in Table 1, where the entries in column 5 are in units of the Hubble time, $H_0^{-1}$, for $H_0 = 50$.

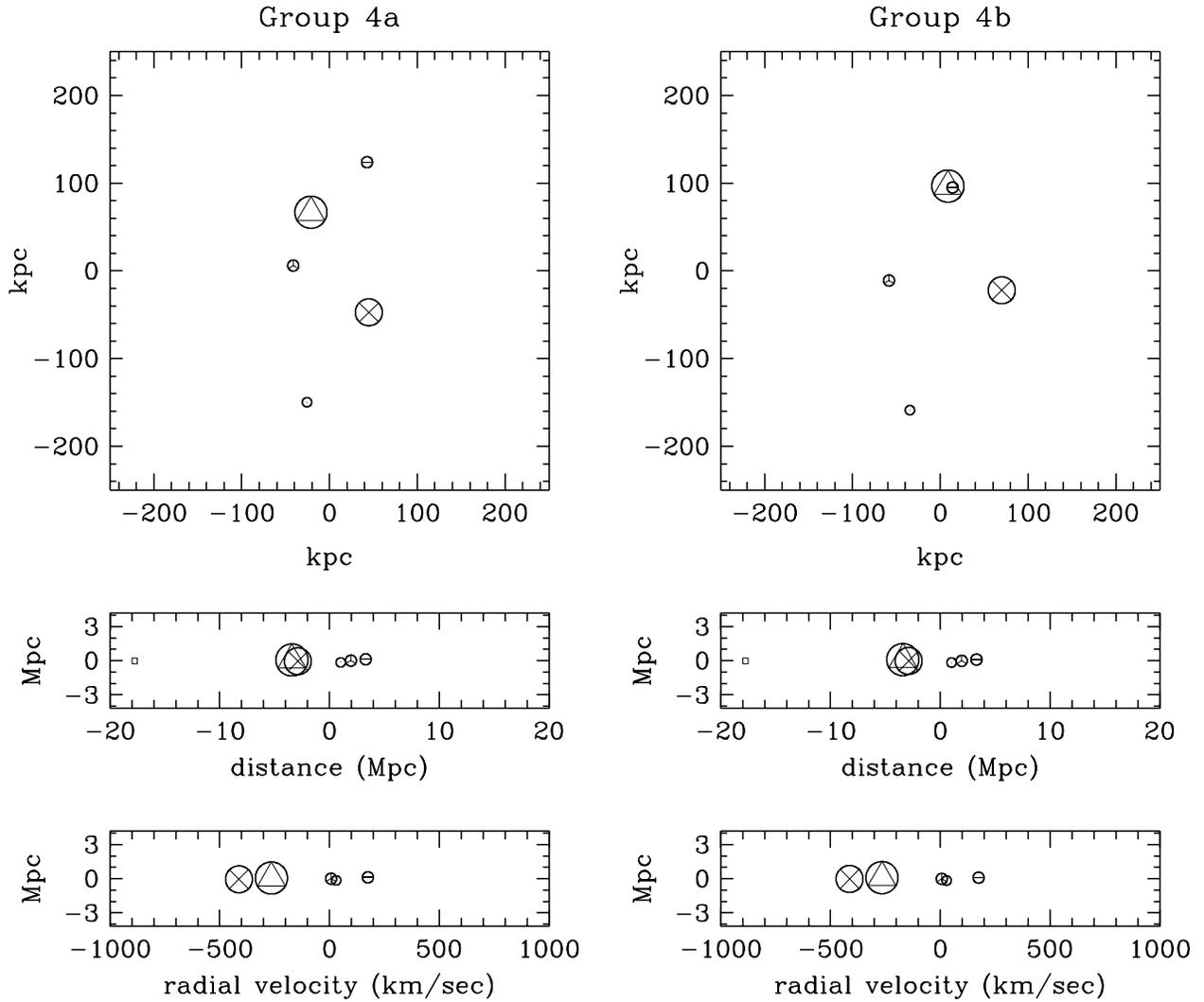

**Figure 1** — continued.

For the groups in Hickson's (1993) catalog, Hickson et al. (1992) obtain a median "size" $<R> = 39h^{-1}$ kpc, or $<R> = 78$ kpc with $h = 0.5$, as in our simulations. The values listed for $R$ in Table 1 are in comfortable agreement with those of observed compact groups. On average, our groups are only slightly larger than the median compact group in Hickson's sample, and none are larger than the most extended of Hickson's groups, for which $R = 270$ kpc. The median radial dispersion found by Hickson et al. (1992) is $<\sigma_z> = 200$ km/sec, in astonishingly good agreement with the values in column 3 of Table 1, aside from Groups 1a and 1b. In this regard, we note that Groups 1a and 1b, although having somewhat extreme $\sigma_z$, are still within the range covered by Hickson's groups, as the largest $\sigma_z$ in his catalog is 617 km/sec. Groups 1a and 1b should probably not be compared directly with observed dynamical properties, however, as these two groups would not have been included in the sample of groups for which Hickson et al. (1992) made dynamical estimates because they contain member galaxies with radial velocities differing by more than 1000 km/sec from the group centroid. Finally, the median crossing time of the groups in the Hickson catalog is $H_0 t_c = 0.016$, in excellent agreement with the estimates in column 5 of Table 1.

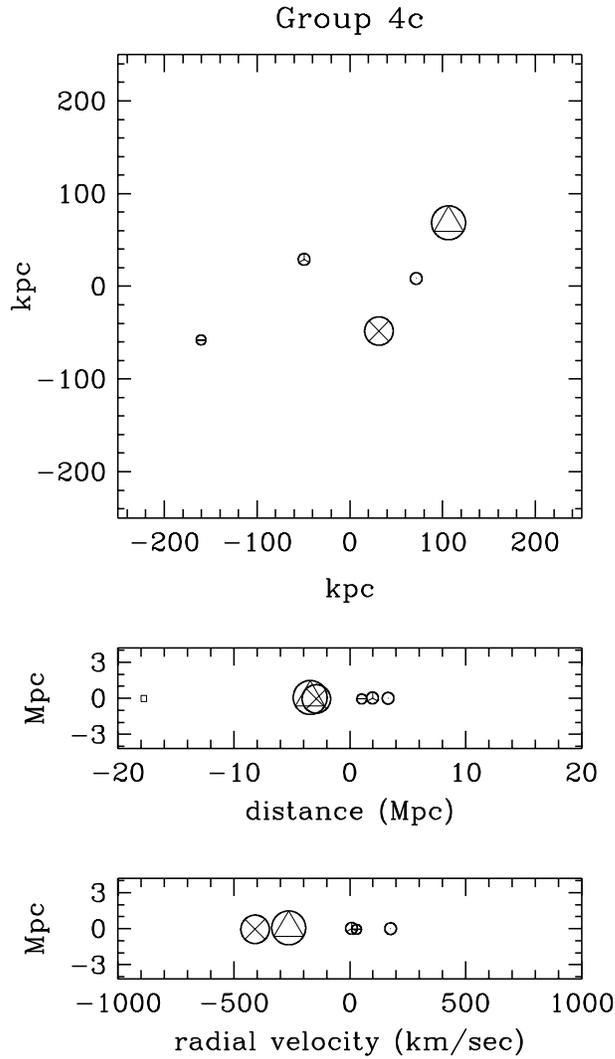

**Figure 1** — continued.

Clearly, the projected groups in Figure 1 have structural and dynamical properties quite similar to the median compact group in the Hickson (1993) sample. However, because they are quite extended along the line of sight, it is equally clear that estimates of merger time scales based on their projected size and kinematics are inappropriate. If, instead of using the median *projected* separation of the galaxies in a simulated group to estimate the crossing time, we characterize the "size" of a group by its median pairwise galaxy separation along the line of sight, denoted by $D_z$ and listed in column 7 in Table 1, we obtain formal "crossing" times for the groups in Figure 1 that are of order a Hubble time. Moreover, since galaxies in filaments are expanding with the Hubble flow, our groups are certainly not virialized and not even collapsing. It is clearly not meaningful to define a "coalescence" time for galaxies in such configurations.

### 4. DISCUSSION

While not definitive, our results raise the interesting possibility that some or even most compact groups are actually detached systems. Unlike earlier suggestions that these groups are chance alignments, ours does not require that the galaxies comprising the "group" be part of a less tightly bound configuration. While we have restricted our investigation to the particular example of structure in a cold dark matter universe with $\Omega = 1$, we believe that this result should be rather model-independent, provided that the galaxy distribution is filamentary. To the extent that this is indeed a property of the observed universe, we expect our proposal to apply, regardless of the precise value of $\Omega$ and the nature of the dark matter. In this regard, our interpretation of compact groups is less model-specific than others, which may require fine-tuning to yield the present collapse rate to account for the observed groups. For example, cosmological models with $\Omega < 1$ may fail in this context, since the rate of merging drops rapidly at low redshifts.

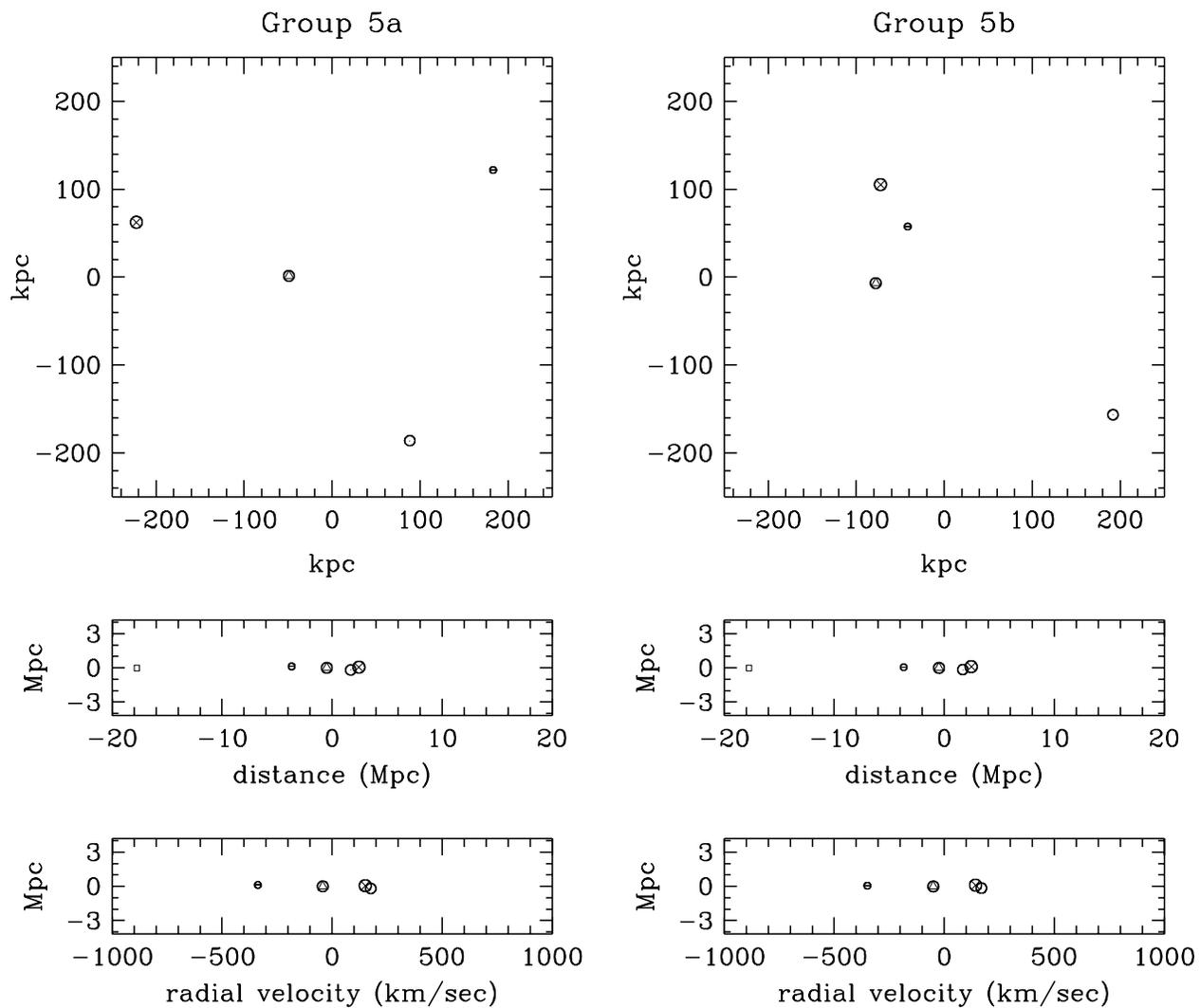

**Figure 1** — continued.

Obviously, an important question for our proposal is whether the probability of viewing filaments end-on is sufficient to account for the observed number of compact groups. Unfortunately, we cannot give a definitive answer to this question because of our small simulation volume, but our best estimate implies that a large fraction of compact groups could arise from such projections. To this end, we examined the filament-projections in Figure 1 to see over what solid angle they would satisfy the selection criteria adopted by Hickson (1993) for groups with at least four accordant members. Groups 1-5 would thus be classified as compact groups over 3.5 × 3.5, 9 × 10, 7 × 7, 9 × 7, and 6 × 5 square degrees, respectively, along one direction of the filaments in which they are contained. To estimate the probability that a random observer in the simulation volume would view one of these filaments as a "compact group", we must double these numbers, since a projection down the opposite side of the filament would yield a similar result. Therefore, the groups in Figure 1 would be called "compact" by Hickson's criteria over a total solid angle of 488.5 square degrees (0.149 steradians), and a random fraction of the cube would yield an apparent compact group $0.149/4\pi = 1.18\%$ of the time. The effective space density of spurious compact groups produced by these projections would be $0.0118/(22 \text{ Mpc})^3 = 1.1 \times 10^{-6} \text{Mpc}^{-3}$, for $H_0 = 50$.

The space density of observed compact groups, $n_{gr}$, is uncertain. From the APM survey and Rose's (1977) catalog, Barnes (1989, 1994) estimates $n_{gr} \sim 1-3 \times 10^{-6}$ Mpc$^{-3}$ for groups with four or more members and $H_0 = 50$. Using a Monte Carlo technique, Mendes de Oliveira & Hickson (1991) obtain $n_{gr} \sim 5 \times 10^{-6}$ Mpc$^{-3}$ for those groups in the Hickson (1993) atlas having four or more accordant galaxies, although this value is quite sensitive to assumptions about the incompleteness of this sample (Barnes 1994). Our estimate of $1.1 \times 10^{-6}$ Mpc$^{-3}$ is at the low end of the plausible range for $n_{gr}$. To some extent, this is to be expected, since the limited resolution in our simulations implies that our galaxy mass function is incomplete at the low-mass end. Because of this, and also because of the small size of our simulation volume, we cannot draw strong conclusions, but our estimates admit the possibility that projected filaments are sufficiently common to produce most of the observed compact groups.

Nevertheless, it must be acknowledged that much circumstantial evidence has accumulated during the past decade to support the notion that compact groups are actually bound systems. In light of the proposal made here, it is useful to reexamine some of the relevant observations. A large fraction ($\sim 30-40\%$) of galaxies in the Hickson compact groups are in some way "peculiar" (Rubin, Hunter & Ford 1991; Zepf & Whitmore 1994; Mendes de Oliveira & Hickson 1994), perhaps indicating that they have experienced interactions and subsequent dynamical evolution. Unfortunately, the significance of these claims is obscured by the lack of a suitable control sample. Moreover, the presence of an interacting *pair* does not demonstrate that the entire compact group is physically bound. Indeed, four of the five projected filaments that make up Figure 1 contain one (but only one) physically close pair. This result is not surprising, since the inclusion of a genuine pair reduces by one the number of background objects needed to satisfy the compact group selection criteria. Only if all the galaxies in a group are *mutually* interacting can it be inferred that they are bound. The Hickson catalog does, in fact, contain several clear-cut examples of groups in this phase of evolution, *e.g.* Hickson 31 and 54. However, such configurations appear to be rare. The absence of larger numbers of compact groups in an advanced stage of merging, and the existence of many groups that have exceedingly short crossing times ($\sim 10^8$ years) but whose galaxies seem relatively quiescent (*e.g.* Hickson 6 and 8), would appear to pose serious difficulties for the hypothesis that the majority of these systems are physically bound.

The existence of intragroup gas (*e.g.* Bahcall, Harris & Rood 1984; Mulchaey *et al.* 1993; Ponman & Bertram 1993) is also cited as evidence that at least some compact groups are actually bound systems. However, simulations like those of Katz *et al.* (1992) show that primordial gas is shock-heated to high temperatures as it converges into filaments. Outside of the high-density cores, the cooling times for this gas are long, so it cannot condense and form stars. When such a filament is viewed end-on, as in Figure 1, it will appear to harbor a compact, bound collection of galaxies embedded in a background of hot, x-ray emitting gas. Thus, the existence of "intragroup" gas does not in itself require that the galaxies in a "compact group" be physically bound. The x-ray properties of a projected filament may differ from the properties of observed physical groups and clusters. Indeed, there is preliminary observational evidence that the ratio of diffuse gas mass to stellar mass is lower in compact groups than in typical loose groups: in a sample of 12 compact groups observed with the ROSAT satellite, only 30%–50% have detectable diffuse x-ray emission, and only Hickson 62 has emission similar to that of a typical loose group, *i.e.* a mass of hot, x-ray emitting gas that is similar to the mass in stars (Pildis, Bregman & Evrard 1993; Bregman 1994).

Studies at radio (Menon & Hickson 1985) and IR wavelengths (Hickson *et al.* 1989) have yielded ambiguous results. Compact groups as a whole do not appear unusually radio-bright (Sulentic 1987), although it has been claimed that some of the galaxies exhibit excess FIR emission, in a manner reminiscent of well-known interacting pairs (for a review, see Barnes & Hernquist 1992). This latter finding is disputed by Sulentic and de Mello Rabaça, who argue that it is an artifact of the limited resolution of the IRAS survey (see also Xu & Sulentic 1991).

It has recently been noted that many compact groups are associated with larger, more loosely bound systems (Ramella *et al.* 1994). This association arises naturally in models where compact groups form within loose groups (Diaferio *et al.* 1994), but it also arises naturally in our filament-projection model for two reasons: the outer parts of the filament can themselves project into a loose background, and real (3-dimensional) groups tend to form at the intersections of filaments, so that an end-on view of a filament is likely to intercept a true loose group.

Our proposal does little to elucidate the presence of galaxies with strongly discordant redshifts in compact groups (*e.g.* Burbidge & Burbidge 1961a; Arp 1973; Sulentic 1983; Sulentic & Arp 1983; Sulentic & Lorre 1983). In particular, the results here neither add to, nor detract from simple estimates that the discordant galaxies are merely interlopers (*e.g.* Hickson, Kindl & Huchra 1988a; White 1990), whose numbers are possibly boosted by gravitational lensing (Hammer & Nottale 1986).

Perhaps the most serious observational challenge to our projection hypothesis is the claim that the morphological mix of galaxies in compact groups differs significantly from that of the field (Hickson, Kindl, & Huchra 1988b; Mendes de Oliveira & Hickson 1994). In particular, it has been argued that the fraction of late-type galaxies in the Hickson compact groups is lower than in the field. However, the significance of this result is unclear owing to ambiguities in classifying S0 galaxies and the fact that the groups were selected from red prints while the field spiral fraction is conventionally estimated from blue prints (Sulentic 1987; Hickson, Kindl & Auman 1989).

More intriguing is the finding that the galaxy types within groups are correlated. Hickson *et al.* (1988b) claim a morphological concordance of galaxies in roughly one-third of the 58 quartets they examined. In the context of our hypothesis, one interpretation of this result is that the spiral-rich groups are projections along filaments while those overabundant in early types are actual bound systems. In that event, it would be natural to expect that the latter subset of compact groups should contain numerous merger candidates; however, this does not seem to be the case. Indeed, there is no evidence of significant ongoing merging in these systems (Hickson *et al.* 1988b; Mendes de Oliveira & Hickson 1991, 1994; Zepf 1993). In addition, there is an underabundance of blue ellipticals relative to the number expected on the basis of evolutionary models (Zepf & Whitmore 1991), and it appears that the luminosity function for the groups as a whole is quite different from that of field ellipticals (Mendes de Oliveira & Hickson 1991, 1994), as should be the case if compact groups eventually coalesce to form these galaxies.

Another possibility is that the morphological mix has been biased as the consequence of observational selection effects. For example, Hickson *et al.* (1988b) have shown that the spiral fraction is lower in the more luminous groups having higher velocity dispersions. However, Mamon (1992) has questioned the significance of this result, noting that group luminosities and dispersions are correlated with redshift, in the sense that the more luminous groups tend to be more distant. Clearly, it is important to determine whether or not the red prints and selection criteria used to identify groups could produce such trends.

## 5. TESTS AND IMPLICATIONS

The real universe doubtless contains both genuine compact groups and apparent compact groups produced by filament projections. The interesting question is whether the filament projections are $\sim 10\%$, $\sim 50\%$, or $\sim 90\%$ of the total. There are three pieces of evidence, two theoretical and one observational, to support our suggestion that projected filaments make up a large fraction of the observed compact groups. First, in our examination of a plausible theoretical model we find that the frequency of projected filaments is similar to the frequency of observed compact groups. This model has not been "tuned" in any way to produce especially filamentary structure; the filaments arise naturally as a consequence of gravitational instability with Gaussian initial conditions. Second, our proposal eliminates long-standing worries over the extremely short dynamical times inferred for many compact groups; these short dynamical times do not imply a merging catastrophe if the "groups" in question are not physically bound. Third, the peculiar morphologies of some compact groups find a natural explanation in the context of our proposal. Many observed compact groups are linear and have a "chained" appearance (Burbidge & Burbidge 1960; Sargent 1968), *e.g.* Groups 14, 27, 29, 34, 39, 49, 54, 55, 66, 72, 88, and 94 of Hickson (1993). At one time interpreted as evidence for explosive ejection from galaxies (Markarian 1961; Arp 1968) or a recent epoch of galaxy formation (Burbidge, Burbidge & Hoyle 1963), such linear structures are, in fact, *expected* when filaments containing galaxies are viewed at slightly oblique angles. This new suggestion is weakly supported by the fact that the "linear" groups listed above have a median redshift $z \approx 0.042$, while the other 80 groups of Hickson (1993) for which reliable velocities are available have a median redshift $z \approx 0.029$. The minimum angle that a filament can make with the line of sight and still be considered a linear group is independent of redshift and depends only on the linearity of the filament and the size of the galaxies within it. The maximum angle that a filament can make with the line of sight and still be considered a compact group is determined by the projected size of the filament, and depends linearly on redshift. Hence, the probability of viewing a linear group increases with redshift, giving linear groups a higher median redshift.

Our suggestion that compact groups are projected filaments admits a clean but challenging observational test: one can search for the true spatial extent of the systems using redshift-independent distance indicators. For our assumed Hubble constant, $H_0 = 50$, the median group in Hickson's sample is at a distance of 174 Mpc, and the typical maximum galaxy separation in our simulated groups is about 7 Mpc. A distance measure that is accurate to at least 4% would be needed to detect the true radial extent of the median group. This accuracy is beyond that of current techniques, with the possible exception of the surface-brightness fluctuation method (Tonry & Schneider 1988; Tonry 1991). However, some compact groups are closer, with at least a half dozen having redshifts of around 0.01 or lower. It might be possible to test our hypothesis with these few groups using existing methods.

Observational support for the projected-filament interpretation of compact groups would have several important implications for our understanding of galaxy formation and galaxy clustering. First, as we have already emphasized, concerns over the dynamical state of compact groups would be eliminated. Second, the notion that many elliptical galaxies could have have originated from the collapse of such groups (*e.g.* Barnes 1989) would have to be reconsidered. Third, virial mass estimates, which have been used extensively to estimate the quantity of dark matter on length-scales characterizing compact groups (Burbidge & Burbidge 1961b; Burbidge & Sargent 1971; Rose & Graham 1979; Hickson *et al.* 1992), would lose their meaning if the groups in question are physically detached. Finally, such observations would provide further evidence for the filamentary character of the distribution of galaxies. Indeed, the statistics of "compact groups" might provide a useful diagnostic for discriminating cosmological scenarios, since models with different fluctuation spectra will yield different predictions for the frequency and properties of galaxy filaments.

In closing, we note that the arguments we have applied to compact groups may also affect the interpretation of binary galaxy statistics (Charlton & Salpeter 1991, Schneider & Salpeter 1992) and the structure and dynamics of loose groups. Clearly it is much easier for a chance projection along a filament to make what appears to be a binary galaxy or a loose group than to make what appears to be a compact group. Equally important, because real groups tend to form at the intersections of filaments, there is a high probability of picking up interlopers from the surrounding structure when observing a genuine group or cluster. Flows converge along the filaments towards these mass concentrations, and this effect can partly cancel the Hubble flow, making it more difficult to weed out interlopers from their redshifts than one might naively expect. Indeed, one could imagine that contamination from a projected filament might either inflate or reduce a group's estimated velocity dispersion, depending on circumstance. If the galaxy distribution has the filamentary character predicted by theory and suggested by observation, then the true spatial distribution of galaxies may be difficult to untangle from the structure seen in redshift space or in projection.

## ACKNOWLEDGEMENTS

We thank Josh Barnes, Mike Bolte, Joel Bregman, Craig Hogan, George Lake, and Tom Quinn for stimulating discussions. This work was supported in part by the Pittsburgh Supercomputing Center and the National Center for Supercomputer Applications (Illinois). LH received support from the Alfred P. Sloan Foundation, NASA Theory Grant NAGW–2422, and from the NSF under Grants AST 90–18526, ASC 93-18185, and the Presidential Faculty Fellows Program. NK received support from a Hubble Fellowship, NASA Theory Grant NAGW–2523, and NASA HPCC/ESS Grant NAG 5–2213. DW received support from the W.M. Keck Foundation and NSF grant AST 92-45317.

TABLE 1

Properties of Projected Groups

| Group | $R$ (kpc) | $\sigma_z$ (km/s) | $t_c$ (yrs) | $H_0 t_c$ | $\Delta z$ (Mpc) | $D_z$ (km/s) |
|---|---|---|---|---|---|---|
| 1a | 128.7 | 606.9 | $1.52 \times 10^8$ | 0.0078 | 20.0 | 243.7 |
| 1b | 66.4 | 607.0 | $7.87 \times 10^7$ | 0.0040 | 20.0 | 243.8 |
| 1c | 62.2 | 183.1 | $2.44 \times 10^8$ | 0.0125 | 4.87 | 216.6 |
| 2a | 138.1 | 95.6 | $1.03 \times 10^9$ | 0.0526 | 6.95 | 168.5 |
| 2b | 143.0 | 96.9 | $1.06 \times 10^9$ | 0.0542 | 6.94 | 168.6 |
| 3a | 96.0 | 230.5 | $3.00 \times 10^8$ | 0.0153 | 7.61 | 123.3 |
| 3b | 171.8 | 232.3 | $5.32 \times 10^8$ | 0.0272 | 7.61 | 123.0 |
| 4a | 124.4 | 214.0 | $4.18 \times 10^8$ | 0.0214 | 6.72 | 113.4 |
| 4b | 128.9 | 213.9 | $4.33 \times 10^8$ | 0.0221 | 6.72 | 113.6 |
| 4c | 122.6 | 212.4 | $4.15 \times 10^8$ | 0.0212 | 6.72 | 113.6 |
| 5a | 261.6 | 205.5 | $9.15 \times 10^8$ | 0.0468 | 6.12 | 146.0 |
| 5b | 112.3 | 207.1 | $3.90 \times 10^8$ | 0.0199 | 6.13 | 146.2 |